\documentclass[]{article}

\usepackage{graphicx}
\usepackage{savesym}
\usepackage{listings}
\usepackage[intlimits]{amsmath}
\savesymbol{iint}
\usepackage{txfonts}
\usepackage{bm}
\restoresymbol{TXF}{iint}
\usepackage{amssymb}
\usepackage[superscript]{cite}
\usepackage{comment}
\usepackage{color}
\usepackage[normalem]{ulem}

\renewcommand{\figurename}{{\bf Fig.}}

\def\apj{Astrophys. J.}
\def\apjl{Astrophys. J. Lett.}
\def\apjs{Astrophys. J. Supp. Ser. }

\def\aap{Astron. Astrophys. }

\def\araa{Ann.\ Rev. Astron. Astroph. }

\def\mnras{Mon. Not. Roy. Astron. Soc. }

\def\pre{Phys. Rev. E.}

\def\be{\begin{equation}}
\def\ee{\end{equation}}
\def\ba{\begin{eqnarray}}
\def\ea{\end{eqnarray}}
\def\go{\mathrel{\raise.3ex\hbox{$>$}\mkern-14mu
             \lower0.6ex\hbox{$\sim$}}}
\def\lo{\mathrel{\raise.3ex\hbox{$<$}\mkern-14mu
             \lower0.6ex\hbox{$\sim$}}}

\begin{document}
\title{Reconfinement and Loss of Stability in Jets from Active Galactic Nuclei}

\author{Konstantinos N. Gourgouliatos\thanks{Department of Applied Mathematics, University of Leeds, Leeds, LS2 9JT, UK}\thanks{
Department of Mathematical Sciences, Durham University, Mountjoy Centre, Stockton Rd, Durham DH1 3LE, UK, Konstantinos.Gourgouliatos@durham.ac.uk}~~ and Serguei S. Komissarov\thanks{Department of Applied Mathematics, University of Leeds, Leeds, LS2 9JT, UK, s.s.komissarov@leeds.ac.uk}}

\date{\today}
\maketitle

{\bf Jets powered by active galactic nuclei appear impressively
stable compared with their terrestrial and laboratory counterpartsÑ
they can be traced from their origin to distances
exceeding their injection radius by up to a billion times\cite{1984RvMP...56..255B, Porth:2015}.
However, some less energetic jets get disrupted and lose their
coherence on the scale of their host galaxy\cite{1984RvMP...56..255B,Laing:2014}. Quite remarkably,
on the same scale, these jets are expected to become
confined by the thermal pressure of the intra-galactic gas\cite{Porth:2015}.
Motivated by these observations, we have started a systematic
study of active galactic nuclei jets undergoing reconfinement
via computer simulations. Here, we show that in the case
of unmagnetized relativistic jets, the reconfinement is accompanied
by the development of an instability and transition to
a turbulent state. During their initial growth, the perturbations
have a highly organized streamwise-oriented structure,
indicating that it is not the Kelvin-Helmholtz instability, the
instability which has been the main focus of the jet stability
studies so far\cite{Birkin91,BMR04}. Instead, it is closely related to the centrifugal
instability\cite{Rayleigh}. This instability is likely to be behind the division
of active galactic nuclei jets into two morphological types in
the Fanaroff--Riley classification\cite{Fanaroff:1974}.} \\

It has been proposed\cite{Porth:2015} that the extraordinary surviving ability of
active galactic nuclei (AGN) jets is due to the fact that outside of the
jet engine they are no longer confined but expand freely in the lowpressure
interstellar environment. This leads to a loss of causal connectivity
across the jet, which suppresses any global instability. Yet,
such a free expansion does not last forever as the thermal pressure of
freely expanding jets rapidly decreases. Eventually, it becomes lower
than the external pressure and the jets become confined again. The
reconfinement involves a strong stationary (or quasi-stationary)
shock that communicates the external pressure to the jet interior.
Once it reaches the jet axis at the reconfinement point (RP), the
process of reconfinement is completed. Downstream of the RP,
causal connectivity across the jet is restored and this is where it may
become vulnerable to global instabilities.

So far, two types of instability have been identified as important
in the dynamics of AGN jets, the KelvinÐHelmholtz instability
(KHI)\cite{Birkin91,BMR04} and the magnetic current-driven instabilities\cite{Bateman78,ALB00}, which have
been explored in detail in numerous studies. Both linear analysis
and numerical studies of jet instabilities have focused on the cylindrical
geometry of the background flow. The geometry of reconfined
jets is more complicated, with curved streamlines even for
non-rotating flows.

Motivated by the idea of causal connectivity, we performed computer
simulations of relativistic jets undergoing reconfinement. As a
first step, only unmagnetised flows were considered. Following the
standard route of stability analysis, we started the investigation of
each particular jet model by constructing its steady-state solution. To this aim, we used the approach from ref.\cite{Komissarov:2015}. This solution
was then projected onto a two-dimensional (2D) and three-dimensional
(3D) computational grid to serve as an initial state for timedependent
simulations.
  
  We considered two different models of the external gas. The
first describes the interstellar gas of elliptical galaxies\cite{MB-03}. It is isothermal
with flat density distribution up to the galactic core radius $r_{c}=1\,$ kpc followed by a decline as $r^{-1.25}$. Reflecting the observations,
this gas is significantly heavier than the jet gas. The second
model is designed to simulate the conditions inside the radio
lobes of FR-II jets. In this model, the external gas is uniform and
significantly lighter than the jet.
  
At the nozzle, the steady-state solution corresponds to a highly
supersonic conical jet emerging from the centre of a galaxy. Its
Lorentz factor ($\Gamma$) is maximum at the axis and decreases towards the
edge of the jet, while its density and pressure are uniform. The key
problem parameter is the jet kinetic power, which determines the
position of the RP\cite{Porth:2015}. We ran several models covering a wide range of
the RP position compared with the galactic core radius. The nozzle
location was selected to allow substantial initial expansion of the jet
before its reconfinement. The size of the computational domain was
at least twice the distance from the nozzle to the RP.

When the gas density and velocity profiles were too steep at
the jet interface, the time-dependent solution was corrupted by
numerical heating, which drove strong shocks into the jet and into
the external gas right at the start of the simulations. To prevent
this from happening, we introduced a boundary layer where the
density changed gradually. With such a modification, our axisymmetric
2D solutions eventually relaxed to a steady state that was
only slightly different from the initial one, indicating its stability.
The stability of such axisymmetric models was reported earlier
by another group\cite{Marti-16}.

In stark contrast with these axisymmetric solutions, our 3D jets
exhibit a completely different evolutionÑtheir reconfinement is
accompanied by a loss of stability and a quick transition to a fully
turbulent state. This is illustrated in Fig.~1, which confronts the 2D
and 3D solutions for a jet propagating through a galactic corona
(model C1 in Supplementary Table 1). One can see that the instability
already develops before the RP in the shocked outer layer of the
still expanding jet, but then it takes hold over the whole of the jet
once the reconfinement shock reaches the jet axis. Such a behaviour
is seen in all our models. The turbulence is very strong and leads
to substantial heating of the jet plasma downstream of the reconfinement
shock. Its pressure rises well above that of the steady-state
model and this drives a shock into the cold interstellar plasma as
well as forcing the RP to move towards the jet origin.

The structure of perturbations before the transition to turbulence
can give away the nature of instability. In our case, the axial slices of the solution do not show the characteristic eddies of KHI, whereas
the transverse jet structure reveals prominent features reminiscent
more of the RayleighÐTaylor instability (RTI). 3D rendering of the
solution shows that these features are in fact stretched along the
streamlines of the background flow (see Fig.~2).

As in rotating fluids, the plasma of reconfined jets moves along
curved streamlines and hence experiences a centrifugal force.
This suggests that our instability could be related to instabilities
found in rotating flows. In particular, it is well known that rotating
flows of an incompressible fluid of uniform density, rotating
with angular velocity $\Omega$, experience the so-called centrifugal instability
(CFI) when the Rayleigh\cite{Rayleigh} discriminant $\Psi=\Omega^{2}R^{4}$  decreases
with distance ($R$) from the centre of rotation. This condition is
always satisfied when the rotation is brought to a halt at a given
distance, which is the case in our jet problem. This instability is
known to produce stream-oriented features like the Taylor vortices
in the Couette flow and the Gšrtler vortices in flows over concave
surfaces\cite{Saric-94}. These vortices spin about the streamlines of the background
flow, which in the context of our problem implies a nonaxisymmetric
motion and hence explains why the instability did
not show up in the 2D simulations. The condition for the development
of the CFI in an axially symmetric relativistic flow, where the assumptions of uniform density and incompressibility have been
relaxed, becomes\cite{GK:2017}
\begin{eqnarray}
\frac{d\ln \Psi}{d\ln R}< M^{2}
\end{eqnarray}
where $\Psi=\Gamma^{2}w\Omega^{2}R^{4}$,  $\Gamma$ is the Lorentz factor,  $w$  is the relativistic
enthalpy and $M=(\Gamma \Omega R)/(\Gamma_{a}a)$ is the relativistic Mach number where
$a$ is the speed of sound and $\Gamma_{a}$ is the corresponding Lorentz factor.
This condition is trivially satisfied in the systems studied here, where
a curved flow is confined by an external medium at rest.

It has been argued\cite{Matsumoto:2013} that the spatial oscillations of reconfined
jets are somewhat similar to the oscillations of initially over-pressured
infinite cylindrical jets with translational symmetry. These
researchers carried out 2D simulations of cylindrical flows and discovered
that they suffered RTI provided the external gas was relatively
light\cite{Matsumoto:2013}. However, their configuration was different. In their
approach, not only the jet but also the external gas participates in
the sideways expansion and the centrifugal force vanishes. In our
simulations, the instability develops independent of the external gas
density, implying that it is not RTI.

Our simulations show that in contrast with idealised cylindrical
flows, in more realistic models of AGN jets one has to consider
the competition between KHI and CFI. The fastest growing KHI
modes in supersonic cylindrical jets are the reflection body modes
for which the jet serves as an acoustic waveguide\cite{PC-85}. The complicated
non-cylindrical structure of reconfined jets may prevent them
from supporting such modes. The growth rate of CFI modes has
to depend on the curvature of the jet streamlines and hence the jet
half-opening angle $\theta_j$. The value of $\theta_j$ used in most of our models
appears a bit high compared with the observations, but $\theta_j$
used in the model C5 is similar to that measured for the jets of
Cygnus A\cite{Bocc-16}. In this model, CFI still dominates over KHI.

AGN jets are magnetized and the magnetic field is believed
to play a key role in their dynamics near the central engine. These
jets are likely to be Poynting-dominated at the start but, even for
the relatively slow ideal collimationÐacceleration mechanism, the
process of conversion of magnetic energy into the kinetic energy
of bulk motion is expected to be completed already at sub-parsec scales\cite{kbvk-07}. Hence, on the kpc scales considered in our study, the
magnetic field is likely to be relatively weak. However, it may still affect
the development of non-magnetic instabilities and introduce new
ones\cite{kbvk-07}. Simulations of magnetized jets is the next step of our study.

The position of the RP point depends on the jet power. For the
typical distribution of the galactic gas, it varies from $\sim50\,$pc for the
weakest AGN jets (of power $L=10^{42}\,$erg s$^{-1}$) to $\sim50\,$kpc for the most
powerful ones ($L=10^{47}\,$erg s$^{-1}$)\cite{Porth:2015}. This agrees with the observed wide
range of flaring distances for AGN jets\cite{Laing:2014}. However, for the high end
of the power distribution, our approach becomes less justified. It is
unlikely that at such large distances the adopted model of the external
gas distribution still holds. Moreover, the longest jets appear to
be shielded from the external medium by light cocoons.

On the large scale, AGN jets inflate bubbles of radio-emitting
plasma, the so-called radio lobes. In the less powerful FR-I sources,
the lobe brightness peaks within the inner half of their extent, whereas in the more powerful FR-II sources, this happens within the outer half, often close to the very edge\cite{Fanaroff:1974}. FR-I jets appear to be
disrupted and dissolve into the lobes. FR-II jets can often be traced
all the way to the most distant parts of the radio lobes, where they
terminate at bright hot spots. Our results explain the FR-I jets as
those destroyed by instabilities triggered when they are reconfined
by the pressure of the intra-galactic gas.

The basic dynamic theory of FR-II jets predicts that they are
reconfined by the bubbleÕs thermal pressure instead. This occurs
well before the jets terminate at the leading hot spots\cite{F-91,Komissarov:1998} and hence
they are expected to experience global instabilities as well. However,
in contrast with FR-I jets, their surrounding is much lighter, which
has important implications for the overall jet evolution. Indeed, our
simulations of jets reconfined by light gas also show rapid onset of
the centrifugal instability at RP, after which the jets still become
turbulent but they do not flare and remain rather well preserved
(see Figs.~3 and 4). Most likely, this reflects the differences in the
turbulent mass entrainment\cite{Bicknell-84,1994MNRAS.269..394K}. When the external gas is light, the
rate of mass entrainment is lower and so is its effect on the jet flow.
This may explain why FR-I jets turn sub-relativistic, whereas the
FR-II jets show signatures of relativistic motion all the way up to the
leading hot spots\cite{Bicknell-84,1994MNRAS.269..394K}.

Many AGN jets can be traced to much smaller distances compared
with the kpc reconfinement scale discussed here. They
usually display a knotty structure, which is often considered as
an indication of an unsteady central engine. Most of these knots
are seen to be moving away from the centre of the parent galaxy.
However, a few of them appear to be more or less stationary and
these have been associated with quasi-stationary shocks\cite{CMA-14,2017ApJ...846...98J}. Since
the jets survive well beyond these shocks, they should differ from
the reconfinement shocks considered here. Instead, they could be
recollimation shocks, which never reach the jet axis, leaving it causally
disconnected and globally stable\cite{KBB12}. Another option involves
non-destructive jet interaction with clouds of interstellar gas or
even stellar winds from clusters of young stars\cite{1994MNRAS.269..394K, 2015MNRAS.447.1001W}.

In a small fraction of AGN, one extended radio lobe has the FR-I
morphology, while another has the FR-II morphology\cite{GW-00}. It has been
argued that the existence of such hybrid sources supports the idea
that the source morphology is not dictated by the specifics of AGN,
which is the same for both lobes, but by the jet interaction with the
external gas, which may have a highly anisotropic distribution. Our
study clarifies the nature of this interaction.

In summary, our results provide strong support to the ideas that:
(1) the long-term stability of AGN jets is due to their rapid expansion;
and (2) the FR-division of radio sources is connected to the
eventual termination of this expansion and onset of instabilities.
Unexpectedly, we also find that the transition from the laminar to
the fully turbulent flow occurs about the PR and it is very sharp
compared with what is normally seen in laboratory supersonic
jets. This sharp transition may explain the observed geometric and
brightness flaring of FR-I jets on the kiloparsec scale\cite{Laing:2014}. According to
the observations, FR-I jets also show significant deceleration at the
flaring point, after which the flow often becomes slower at its edges
than it is on-axis. This is exactly what occurs in our 3D models.

\begin{figure}
\includegraphics[width=0.985\columnwidth]{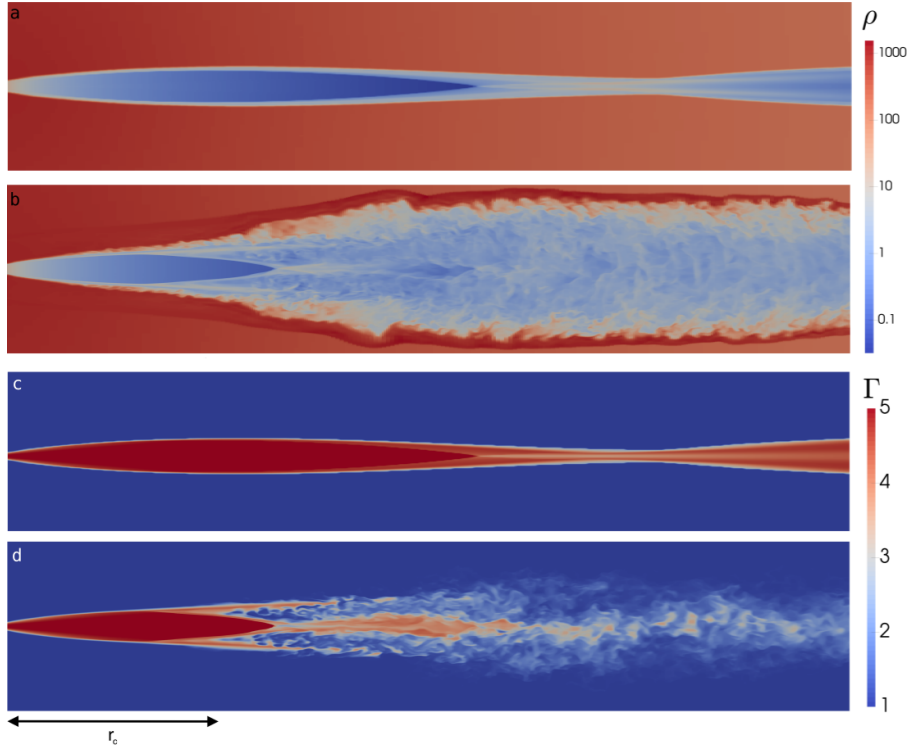}
\caption{ {\bf Jet density ($\rho$, panels a and b) and Lorentz factor ($\Gamma$ panels c and d), for the model C1, corresponding to an FR-I jet making its way through a galactic corona.}
{\bf a--d}, The $y = 0$ section of the jet, containing the axis of the jet, is shown
for $\rho$ ({\bf a}) and $\Gamma$ ({\bf c}) in the steady state solution and $\rho$ ({\bf b}) and $\Gamma$ ({\bf d}) in the
final 3D solution at $t = 32.6$~kyr. The origin of coordinates is located at the
galactic centre and the z axis runs along the jet. The radius of the core
of the galaxy is $r_{c}=1\,$kpc. The density is expressed in arbitrary units and
this applies for all subsequent figures (please refer to Supplementary
Table 1 where the details of the run are presented). When analysing such
plots, one has to remember that the 3D flows are no longer axisymmetric
and significant structure may exist in the azimuthal direction. The 3D
solution has a recessed RP compared with the steady-state one, which
is a general property of all our models. Downstream of the RP, the
flow is highly turbulent. The Lorentz factor distribution looks narrower
compared with the density distribution, indicating that the jet has
developed a fast-spine-slow-sheath structure in the turbulent zone. The
dense shell surrounding the jet is a shock-compressed external gas. The
dissipation of jet kinetic energy in the turbulent zone causes its heating
and sideways expansion, which drives a quasi-cylindrical shock through
the cold gas of the galactic corona. Longer runs are required to reach the
stationary in the statistical sense phase that must follow this
initial expansion.}
\end{figure}

\vskip 0.5cm

\begin{figure}
\includegraphics[width=0.985\columnwidth]{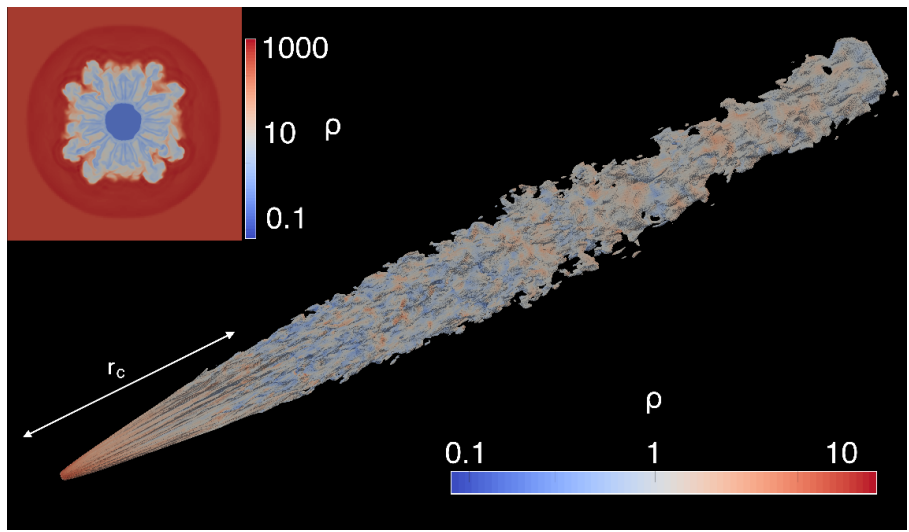}
\caption{ {\bf 3D rendering of model C1 at the end of the run t = 32.6 kyr.}
The insert shows the corresponding density distribution in the cross-section
$z\approx r_c$, just before the flow turns turbulent. Based on this inserted
image alone, one may conclude that we are dealing with the RTI, but
the 3D rendering shows that the perturbations are stretched along the
streamlines of the background flow and form a rather regular pattern in
the first quarter of the jet length. Such patterns have never been seen
before in laboratory jets or jet simulations, but are known in other flows.
The most familiar examples are the Taylor and G\"ortler vortices, which are
associated with the CFI.}
\end{figure}
\vskip 0.5cm

\begin{figure}
\includegraphics[width=0.985\columnwidth]{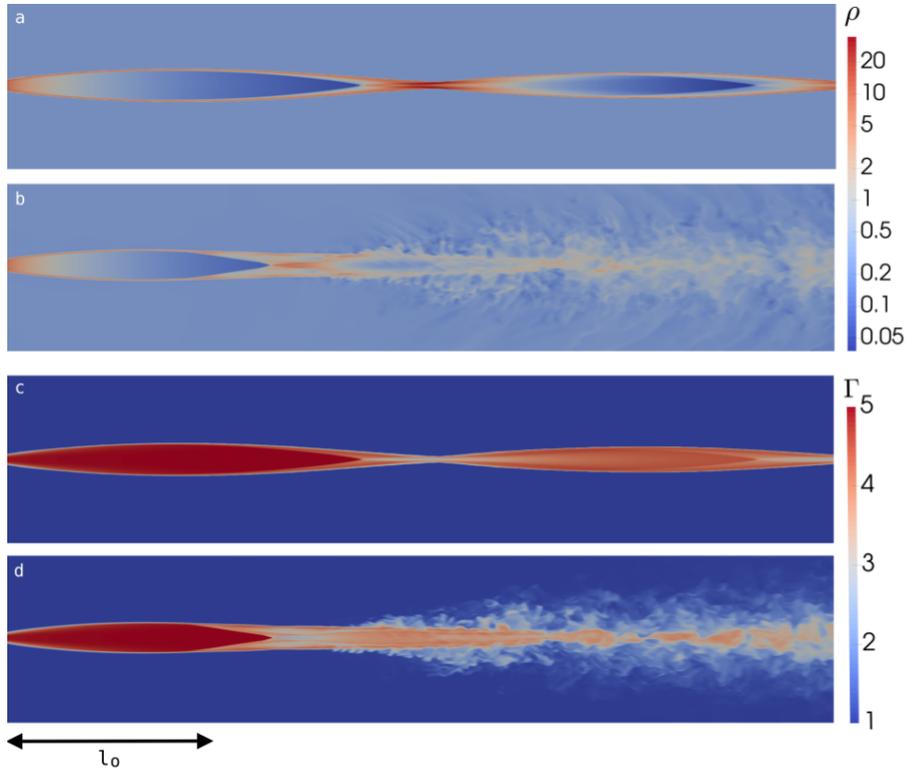}\\
\caption{{\bf  Jet density ($\rho$) and Lorentz factor ($\Gamma$) for the model U2,
representative of an FR-II jet. a -- d,} Solutions at the beginning of the run
for $\rho$ ({\bf a}) and $\rho$ ({\bf c}) and at the end of the run $\rho$ ({\bf b}) and $\Gamma$ ({\bf d} ) ($t = 32.6$kyr),
with $l_{0}=1\,$kpc, for a section as in Fig. 1. Similar to the C1 case, the jet
turns turbulent on the reconfinement scale but it appears less disrupted,
with a fast spine surviving all the way up to the right boundary of the
computational domain.}
\end{figure}
\vskip 0.5cm

\begin{figure}
\includegraphics[width=0.985\columnwidth]{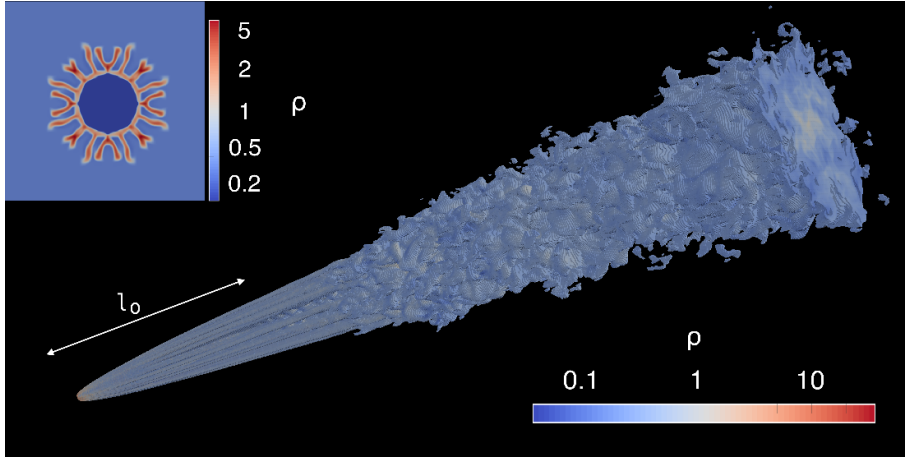}
\caption{{\bf 3D rendering of model U2 at the end of the run t = 32.6 kyr.}
The insert shows the corresponding density distribution in the cross-section
 $z\approx l_0$, just before the flow turns turbulent. The jet appears to be
wider compared with the C1 jet in Fig. 2, but this is not significant and
conveys more about the properties of the numerical diffusion than the
actual jet dynamics. }
\end{figure}

\vskip 0.5cm

\newpage

\vskip 1cm


\section*{Methods}

\noindent 
Below, we present the setup used for the simulations described in this study.
Throughout this work, we used the AMRVAC code as described in refs\cite{Keppens:2012,Porth:2014},
integrating the equations of relativistic hydrodynamics
\begin{eqnarray} 
&& \partial_t \left(\rho\Gamma\right) + \nabla_i \left(\rho \Gamma v^i\right) = 0\,,  \label{eq:cont} \\ 
&& \partial_t \left(\rho h \Gamma^2 - p\right) + \nabla_i \left(\rho h \Gamma^2 v^i\right) = S_0 \,, \label{eq:enth} \\ 
&& \partial_t \left(\rho h \Gamma^2 v_j\right) + \nabla_i \left( \rho h \Gamma^2 v^i v_j + p \delta^i_j \right) = S_j \,,
\label{eq:mom}
\end{eqnarray} 
where  $\Gamma$  is the Lorentz factor, $v$ is the velocity, $\rho$ is the rest mass density in the
fluid frame, $p$ is the pressure and $h= 1+\gamma/(\gamma-1) \left(p/\rho\right)$ is the specific enthalpy
of ideal gas. $S_0$ and $S_j$ are the source terms introduced to keep the external gas in
equilibrium (see below). In the simulations, $\gamma=4/3$. We used the HLLC Riemann
solver\cite{Harten:1997}, Koren flux limiter and two-step RungeÐKutta method for time integration.
To implement this, we used the built-in routines of the code.
 
\vskip 0.5cm
\noindent {\bf Steady-state Solutions.}  Steady-state solutions. To study the stability of a steady-state solution describing
a jet reconfined by the thermal pressure of the external gas, we needed to find
this solution first. To this aim, we used the approach described in ref. \cite{Komissarov:2015}, where an
approximate steady-state solution for an axisymmetric jet was found by solving a
corresponding one-dimensional time-dependent problem in cylindrical geometry.
Such solutions are sufficiently accurate provided the jet is narrow and highly
relativistic, with the jet speed $v \approx c$. In these time-dependent simulations, the initial
conditions describe the parameters at the nozzle of the steady-state problem. Here,
we assume the initial jet Lorentz factor
\begin{eqnarray} \Gamma= \left\{
\begin{array}{ll}
\Gamma_0\left[1-\left(\frac{R}{R_{j}}\right)^4\right]+\left(\frac{R}{R_{j}}\right)^4\,,
& R\leq R_{j} \\ 
1 \,, & R> R_{j}\, \\
\end{array} \right.
\end{eqnarray}
where $R$ is the cylindrical radius, and the velocity vector
\begin{equation} 
(v_{R}, v_{\phi}, v_z) = v\, (\sin\theta,0,\cos\theta)\,
\end{equation}
where $v$ is the speed corresponding to $\Gamma$, $\theta=\arctan\left(R/z_{0}\right)$ and $z_0$ is the nozzleÕs
distance from the jet origin. The $z$ direction is aligned with the jet axis and hence
this velocity distribution corresponds to a conical jet emerging from the origin.
In the simulations, we assumed that $z_{0}=0.1 r_c$ and $R_{j}=0.02r_c$, where $r_c$ is the
galactic core radius. These give the half-opening angle $\theta_j =0.2$. We also carried
out a simulation with an opening angle of $\theta=0.1$, $\Gamma_{0}=5$ and $L_{j,44}=2$ (see run
C5 in Supplementary Table 1). Although the opening angles of AGN jets are
somewhat smaller, the position of the RP only weakly depends on the opening
angle for as long as $\theta_j\ll 1$. Our choice of $\theta_j$ is dictated by the need to keep the
computational time of 3D simulations manageable. Both the initial density and
pressure distributions are uniform across the nozzle. The steady-state solution is
reconstructed from the time-dependent one via the substitution $z=z_0+ct$.. We
consider an external pressure that depends on the distance from the origin so that
$p_e=p_e(r)$ and we use the substitution:
$$
r(t)^2=R^2+z(t)^2\,.
$$
so that the pressure becomes a function of the cylindrical distance and time. 

We identify the external gas using the level set method\cite{SS-03} with a passive tracer $\tau$
satisfying the equation:
\begin{eqnarray} \frac{\partial \left( \Gamma \rho
\tau\right)}{\partial t} + \nabla_i \left(\Gamma \rho \tau 
v^i \right)=0\,.
\end{eqnarray}
The tracer is initialised via
\begin{eqnarray}
\tau=\frac{1}{2}\left[1-\tanh\left(\frac{R-R_{j}}{w}\right)\right]\,,
\end{eqnarray}
with $w=0.05R_j$. The transition from the jet to the external medium occurs when
the passive tracer value drops below the cut-off value $\tau_{b}=0.5$. As we integrate
forward in time, we reset the value of the pressure to $p_e(r(t))$  for every time step for
cells with $\tau <\tau_{b}$.

Coronas of elliptical galaxies are modelled as hydrostatic isothermal
atmospheres with the density distribution

Coronas of elliptical galaxies are modelled as hydrostatic isothermal  atmospheres with the density distribution
\begin{eqnarray} \rho_{e}(r)=\rho_{e,0}
\left[1+\frac{r^2}{r_c^2}\right]^{-a/2}\,,
\label{DENS}
\end{eqnarray} 
where $r_{c}$ is the radius of the galactic core. For all models, we use $r_{c}=1$~kpc and
$a=1.25$--the typical values for giant elliptical galaxies. Since the initial opening
angle of our simulated jets is higher than that of AGN jets, the jet density is lower.
Because the jet-to-external-gas-density ratio is considered as one of the key
parameters in the jet dynamics, we opted to lower the external gas density by the
same factor. Since the position of the RP is determined by the external pressure\cite{Porth:2015},
the latter was kept unchanged with the core pressure $p_{e,0} = 3\times 10^{-10}\mbox{dyn}\,\mbox{cm}^{-2}$,
leading to a higher external temperature. The parameters of jet models covered in
this study are given in Supplementary Table 1.

As a guide, we use the analytical model\cite{Porth:2015} to estimate the position of the RP and
hence decide on the size of the computational domain and the integration time. At
the left boundary ($R=0$), the boundary conditions are dictated by the geometry, which is symmetric for the scalar quantities (density, pressure and passive tracer)
and the axial component of the velocity and antisymmetric for the radial and
azimuthal velocity component. At the right boundary, we impose the zero-gradient
conditions. Our convergency study shows that at the nozzle the jet radius has to be
resolved by at least 20 cells.

\vskip 0.5cm
\noindent {\bf 2D Simulations.}  The 2D simulations are intended to test the jet stability to
axisymmetric perturbations. The size of the cylindrical computational domain
$(0,R_{d})\times(z_0,z_{1})$ is decided based on the parameters of the steady-state solution.
The initial solution is set via projecting the steady-state solution onto the 2D
cylindrical grid. At $R=0$, the boundary conditions are dictated by axisymmetry.
At the external boundary $R=R_{d}$, we use the zero-gradient conditions. At $z=z_0$,
the parameters of the ghost cells are fixed to those of the external gas for $R>R_j$,
whereas for $R<R_j$, we use the parameters of the steady-state jet instead. Finally, the
zero gradient conditions are employed at $z=z_{1}$. In the simulations, we use a single
uniform grid with a resolution of 20 cells per jet radius at the nozzle. The cell shape
is $\Delta R= 0.2 \Delta z$. This is exactly the same shape and resolution as those of the finest
grid in our 3D runs.

Preliminary test runs with such a setup revealed very strong numerical
dissipation at the jet boundary leading to rapid corruption of the initial solution.
We have found that this can be prevented by replacing the density discontinuity at
the jet boundary with a smooth transition layer with a $\tanh$-profile of thickness
$\Delta R=0.1R_{j}$. The same approach is used in the 3D simulations.

When the external gas distribution corresponds to that of galactic coronas, we
need to make sure that it does not evolve unless it is perturbed by the jet. To this
aim, we have implemented the Newtonian gravity model\cite{PM-07}. This is achieved using
the source term
\begin{equation} 
S_{0}=a H \rho \frac{ v^{i}r_{i}}{r^{2}_{c}+r^{2}}
\end{equation} 
into the energy equation and 
\begin{equation} 
S_{j}=a H\rho \frac{ r_{j}}{r^{2}_{c}+ r^{2}}
\end{equation}
into the momentum equation, where $H=p_{e,0}/\rho_{e,0}$.

\vskip 0.5cm
\noindent {\bf 3D Simulation.}  In our 3D simulations, we use the Cartesian domain $(-R_d,R_d)\times(-R_d,R_d)\times(z_0,z_{1})$. $R_{d}$ is set to a few times the largest jet radius in the steady-state
solution and $z_{1}$ to about twice the distance to RP in this solution. The initial 3D
solution is prepared in the same fashion as in the 2D simulations described above.
At $z=z_0$ and $z=z_{1}$ we use the same boundary conditions as in the 2D simulations
and at the remaining boundaries we enforce the zero-gradient conditions. As in the
2D simulations, Newtonian gravity is introduced to keep the unperturbed external
gas in hydrostatic equilibrium.

As a rule, we perturb the initial solution by multiplying the jet density by
$1+10^{-2}\cos\phi$ and its pressure by $1+10^{-2}\sin\phi$, where $1+10^{-2}\sin\phi$. If such
perturbations are not introduced, the growing perturbations are dominated by the
mode with the azimuthal number $m=4$, which is aligned with the Cartresian grid,
indicating that the perturbations come from the discretisation errors. However, the
overall evolution does not change much.

In the 3D runs, we benefit from the adaptive mesh capabilities of the AMRVAC
code. We introduced four levels of adaptive mesh refinement and used the Lohner
refinement criterion\cite{Lohner:1987} based on the behaviour of the Lorentz factor. The resolution
of the base level is set to $100^{3}$, corresponding to 20 cells per nozzle radius at the
finest grid. The cell shape is $\Delta x= \Delta y =0.2 \Delta z$; we have experimented with various
aspect ratios and conclude that for this choice the results are sufficiently close to
those with the aspect ratio of unity.

\vskip 0.5cm
\noindent {\bf Data Availability Statement}

The data that support the plots within the paper and other findings are available from the corresponding author up reasonable request.

Received: 9 June 2017; Accepted: 10 November 2017;

\vskip 0.5cm
\noindent {\bf Correspondence}

Correspondence should be addressed to K.N.G. or S.S.K.

\vskip 0.5cm
\noindent {\bf Acknowledgements}

The authors acknowledge STFC grant ST/N000676/1. Simulations were
performed on the STFC-funded DiRAC I UKMHD Science Consortia machine,
hosted as part of and enabled through the ARC HPC resources and
support team at the University of Leeds. We thank Dr Oliver Porth for
insightful discussions of the intricacies of AMRVAC code. 

\vskip 0.5cm
\noindent {\bf Contributions}

Both authors contributed to planning this research and the analysis of its results. 
All the simulations were carried out by KNG.

\vskip 0.5cm
\noindent {\bf Competing financial interests}

The authors declare no competing financial interests.
\vskip 0.5cm

\newpage

\renewcommand{\figurename}{{\bf Supplementary Figure}}
\renewcommand{\tablename}{{\bf Supplementary Table}}

\setcounter{figure}{0}

\begin{table}
\centering
\begin{tabular}{rccccrcr} \hline\hline
Model & $L_{j,44}$ &$\Gamma_0$ & $\theta_j$ & $\rho_{j,0}/\rho_{e,0}$ & $z_{RP}$ & $x,y,z$  & t \\ 
 \hline
U1 & 2   &  5 & 0.2 & 1                 & 2.1  &  $1.2\times1.2\times4$   & 33  \\
U2 & 2   &  5 & 0.2 & 64                & 2.1  &  $1.2\times1.2\times4$   & 33 \\
C1 & 2   &  5 & 0.2 & $1.6\times10^{-2}$ & 2.3  &  $0.8\times0.8\times4$   & 33  \\
C2 & 2   & 20 & 0.2 & $10^{-3}$          & 2.3  &  $0.8\times0.8\times4$   & 33  \\
C3 & 5   & 10 & 0.2 & $10^{-2}$          & 6.4  &  $1.2\times1.2\times10$  & 65 \\
C4 & 20  & 20 & 0.2 & $10^{-2}$          & 30.0  &  $6.0\times6.0\times40$  &150  \\
C5 & 2   & 5  & 0.1 & $6.4\times10^{-2}$ & 2.2  &  $0.4\times0.4\times4$   & 13  \\
\hline  
\end{tabular}
\caption{{\bf Simulation models. }
The first column is the model name. For cases where the external gas is uniform it starts with letter U and for cases where it describes galactic corona with letter C.  
The second column is the jet power in units of $10^{44}$erg~s$^{-1}$. 
The third is the jet Lorentz factor at the centre of the nozzle. 
The fourth is the initial jet opening angle in radians. 
The fifth is the jet-external gas density ratio at the nozzle. 
The sixth is the  distance of the reconfinement point from the origin in kpc.  
The seventh is the size of the 3D computational domain in kpc.  
The last column column gives the integration time of the 3D runs in kyr.}
\label{Table:1}
\end{table}

\begin{figure}
\includegraphics[width=1.0\columnwidth]{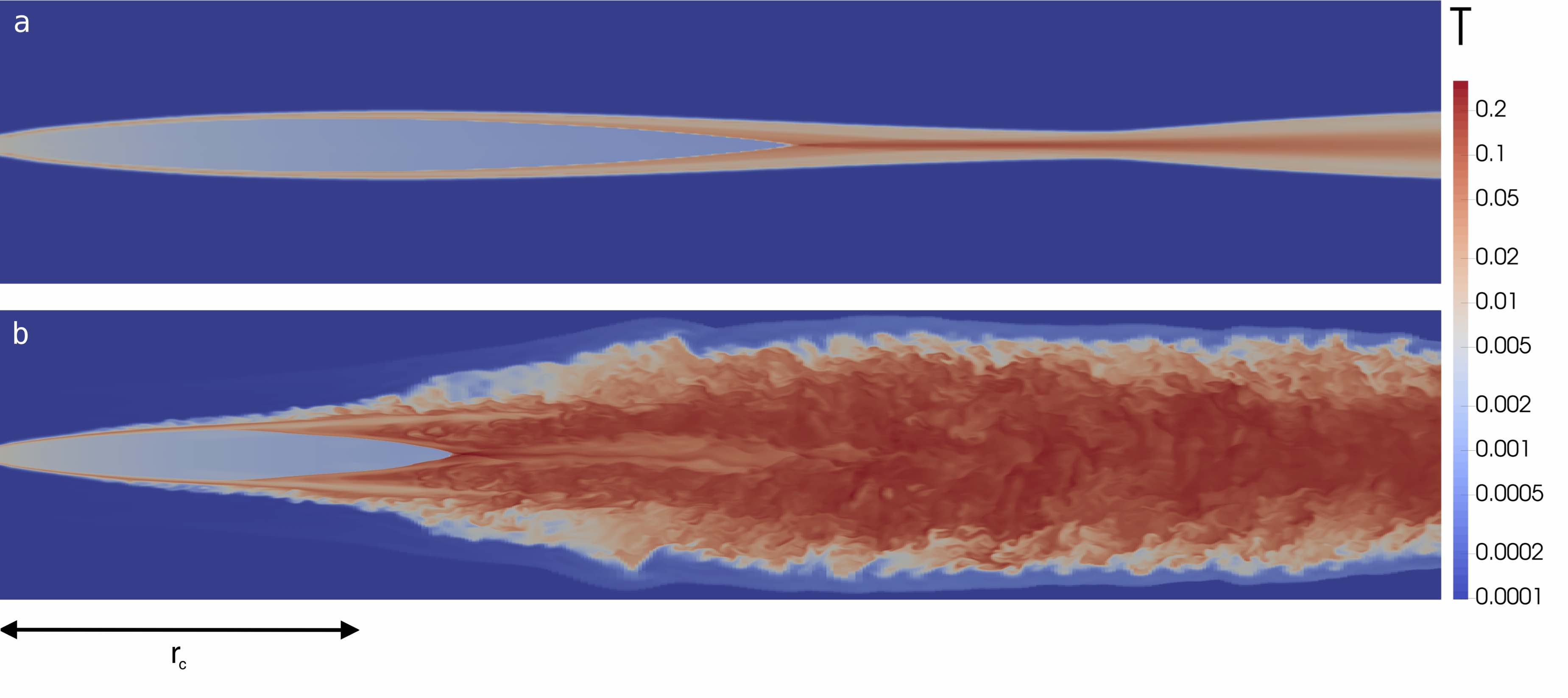}\\
\caption{{\bf  Temperature distribution in the xz-plane of the model C1}. The initial state is shown in panel a; and the final in panel b ($t=32.6\,$kyr). The sharp rise in temperature demonstrates the conversion of the kinetic energy into heat. The units are arbitrary as in Supplementary Figures 2 and 5 and 7. }
\label{Figure:1}
\end{figure}
\begin{figure}
\includegraphics[width=1.0\columnwidth]{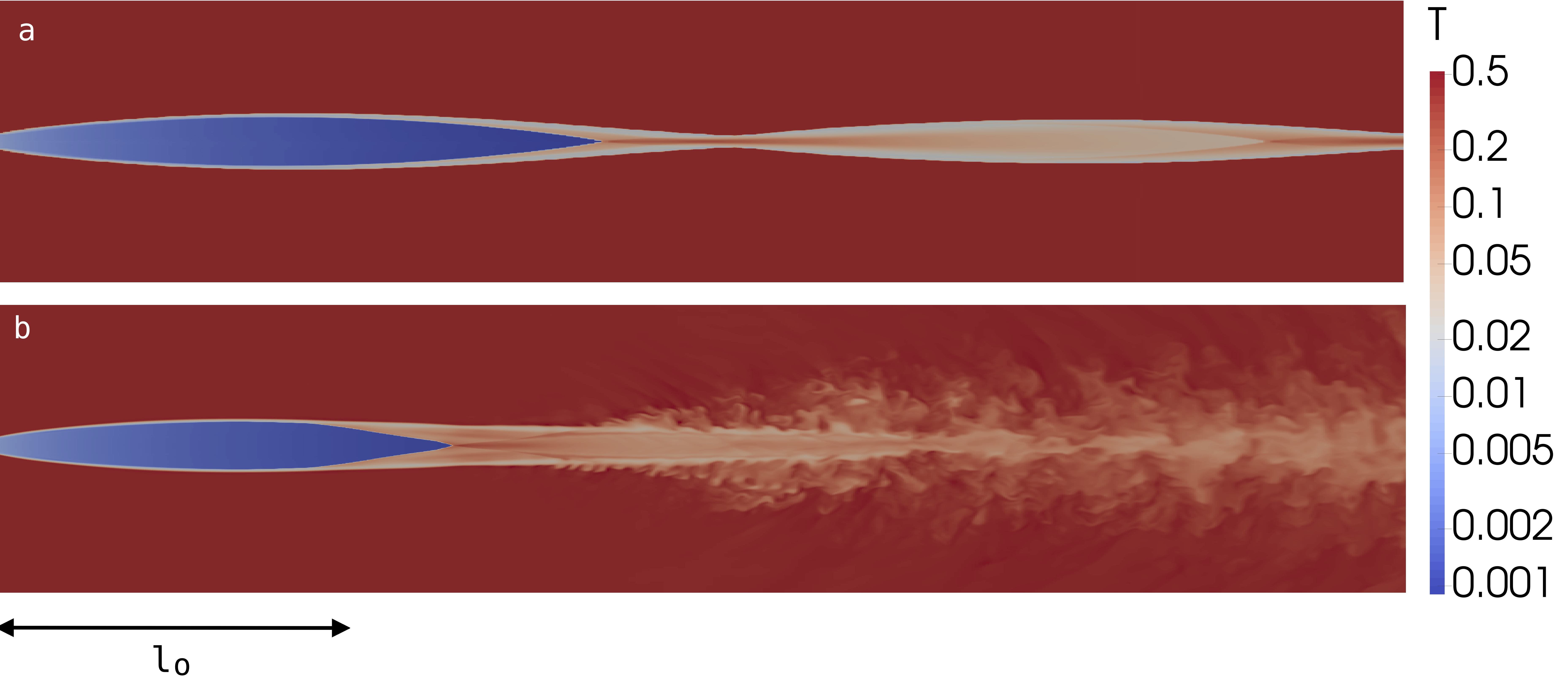}\\
\caption{{\bf Temperature distribution in the xz-plane of the model U2.} The initial state is shown in panel a; and the end of the run in panel b ($t=32.6\,$kyr).  In this model the external gas is lighter compared to the jet, thus representing the case of a jet reconfined by the thermal pressure of its radio lobe. Once the jet becomes unstable, its temperature rises but does not exceed that of the external medium.
}
\label{Figure:2}
\end{figure}
\begin{figure}
\includegraphics[width=1.0\columnwidth]{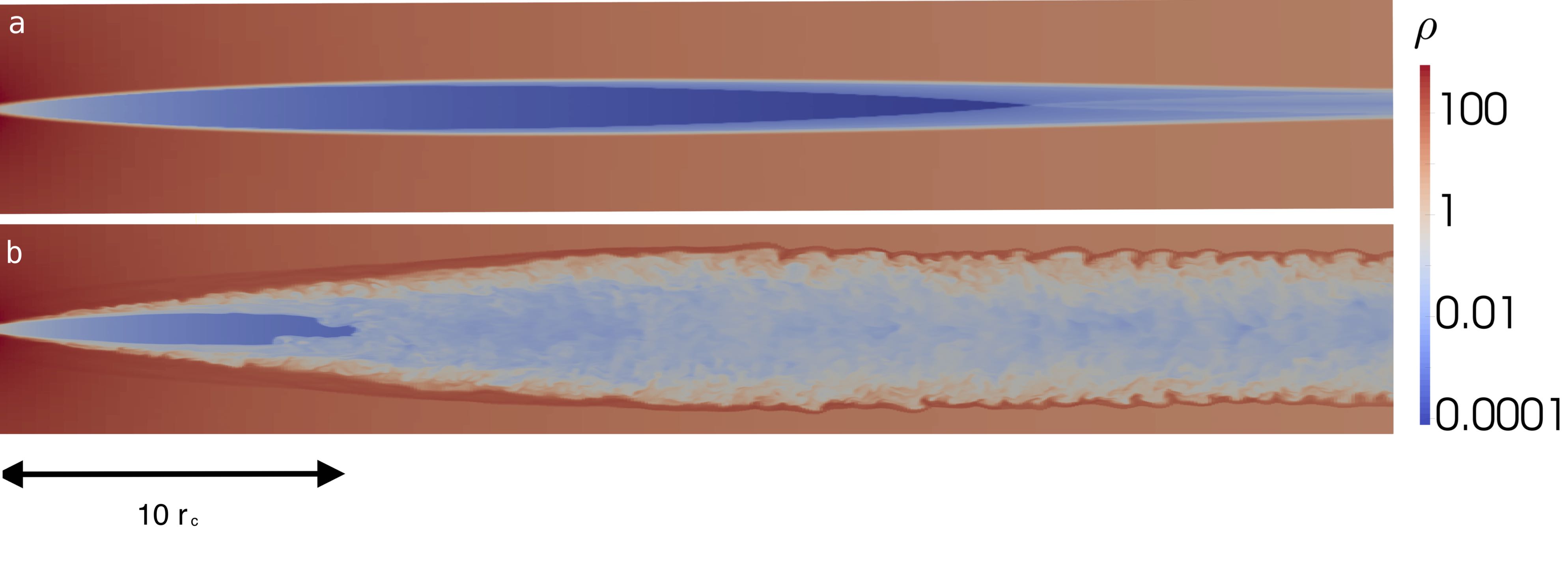}\\
\caption{ {\bf Density distribution in the xz-plane of the model C4}. The initial state is shown in panel a; and the state at the end of the run in panel b ($t=150\,$kyr). This jet is much more powerful compared to the C1 jet and reconfines well outside of the galactic core $r_{c}$. Nevertheless it shows the same behaviour: it develops instabilities and turns turbulent on the reconfinement scale. }
\label{Figure:3}
\end{figure}
\begin{figure}
\includegraphics[width=1.0\columnwidth]{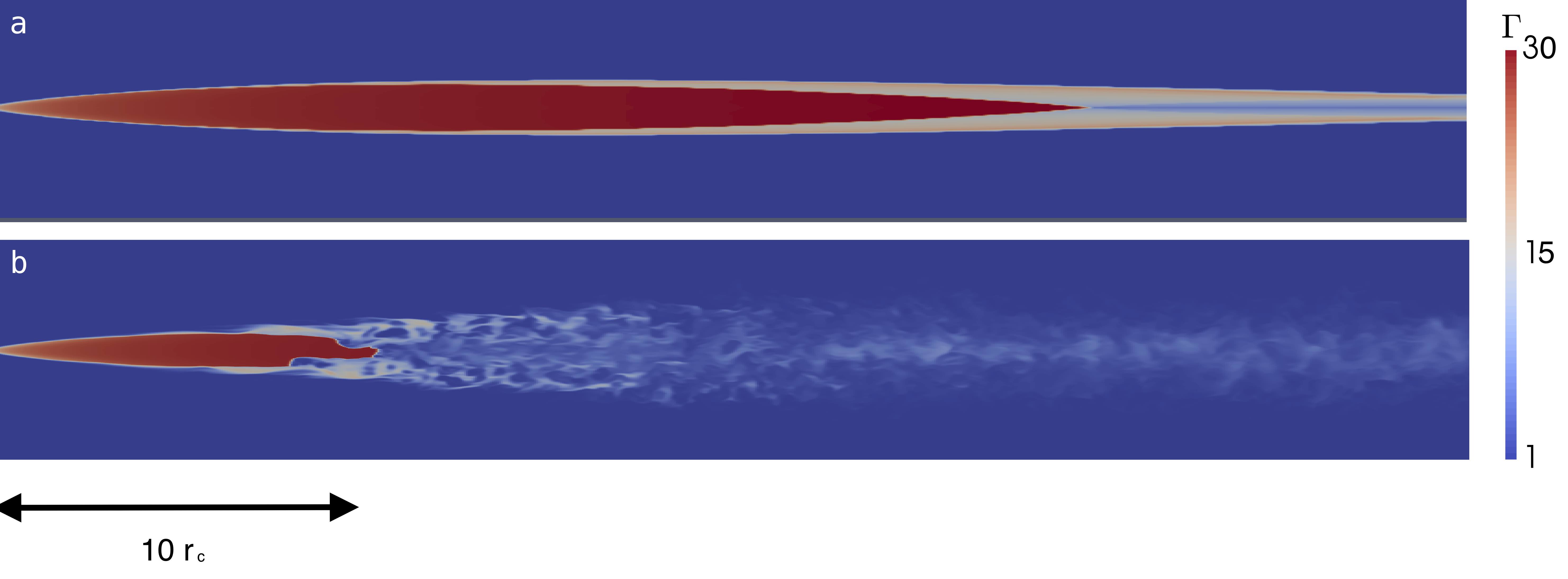}\\
\caption{{\bf Lorentz factor distribution in the xz-plane of the model C4.} The initial state is shown in panel a; and at the end of the run in panel b ($t=150\,$kyr). The very high Lorentz factor of the jet does not prevent it from developing instabilities on the reconfinement scale.  }
\label{Figure:4}
\end{figure}
\begin{figure}
\includegraphics[width=1.0\columnwidth]{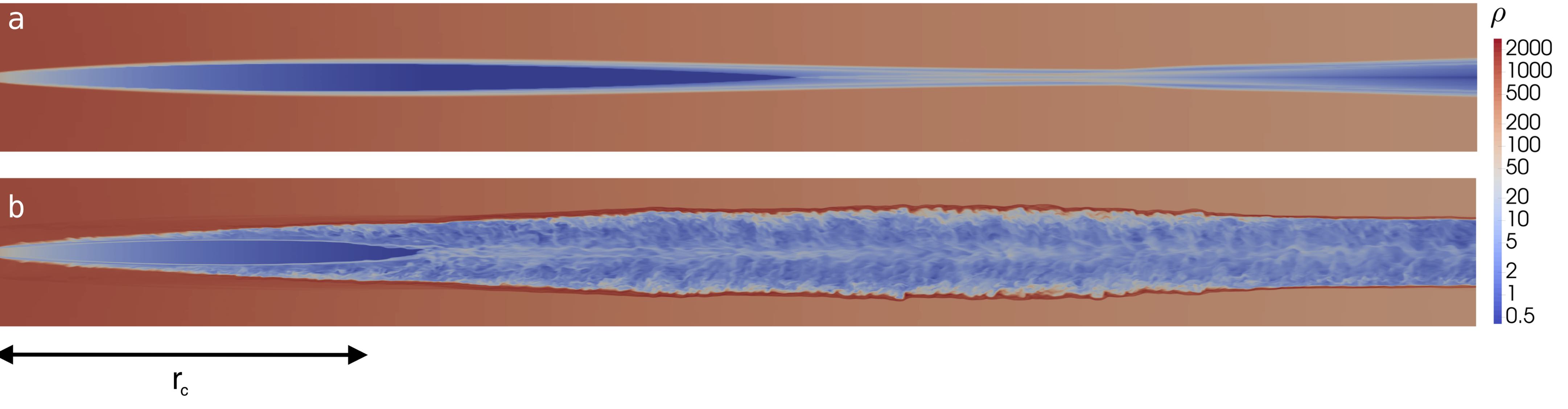}\\
\caption{ {\bf Density distribution in the xz-plane of the model C5.} The initial state is shown in panel a; and at the end of the run in panel b ($t=13\,$kyr), one light-crossing time of the computational domain along the jet axis. This jet has the same power as in the model C1 but its opening angle is smaller by a factor of two, $\theta_j=0.1$rad. Despite the smaller curvature of its streamlines the jet becomes unstable in a similar manner to the other runs.  
}
\label{Figure:5}
\end{figure}
\begin{figure}
\includegraphics[width=1.0\columnwidth]{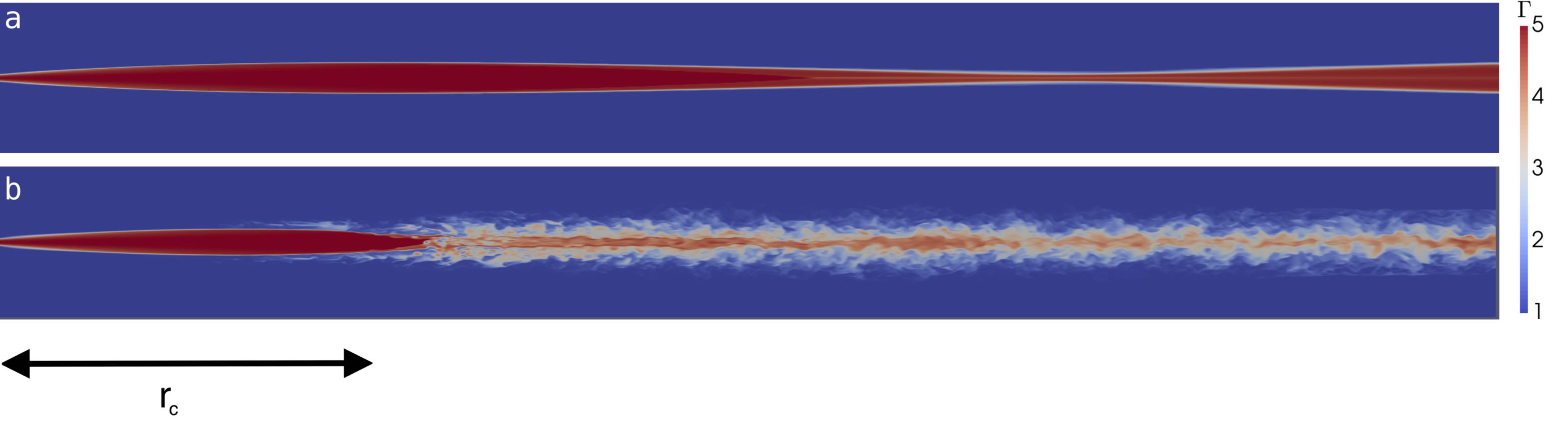}\\
\caption{{\bf Lorentz factor distribution in the xz-plane of the model C5.} The initial state is shown in panel a; and at the end of the run at panel b 
($t=13\,$kyr). Following the destabilisation at PR, the Lorentz factor drops in accordance with the other models. The spine-sheath structure of the jet is manifest as well.  
}
\label{Figure:6}
\end{figure}
\begin{figure}
\includegraphics[width=1.0\columnwidth]{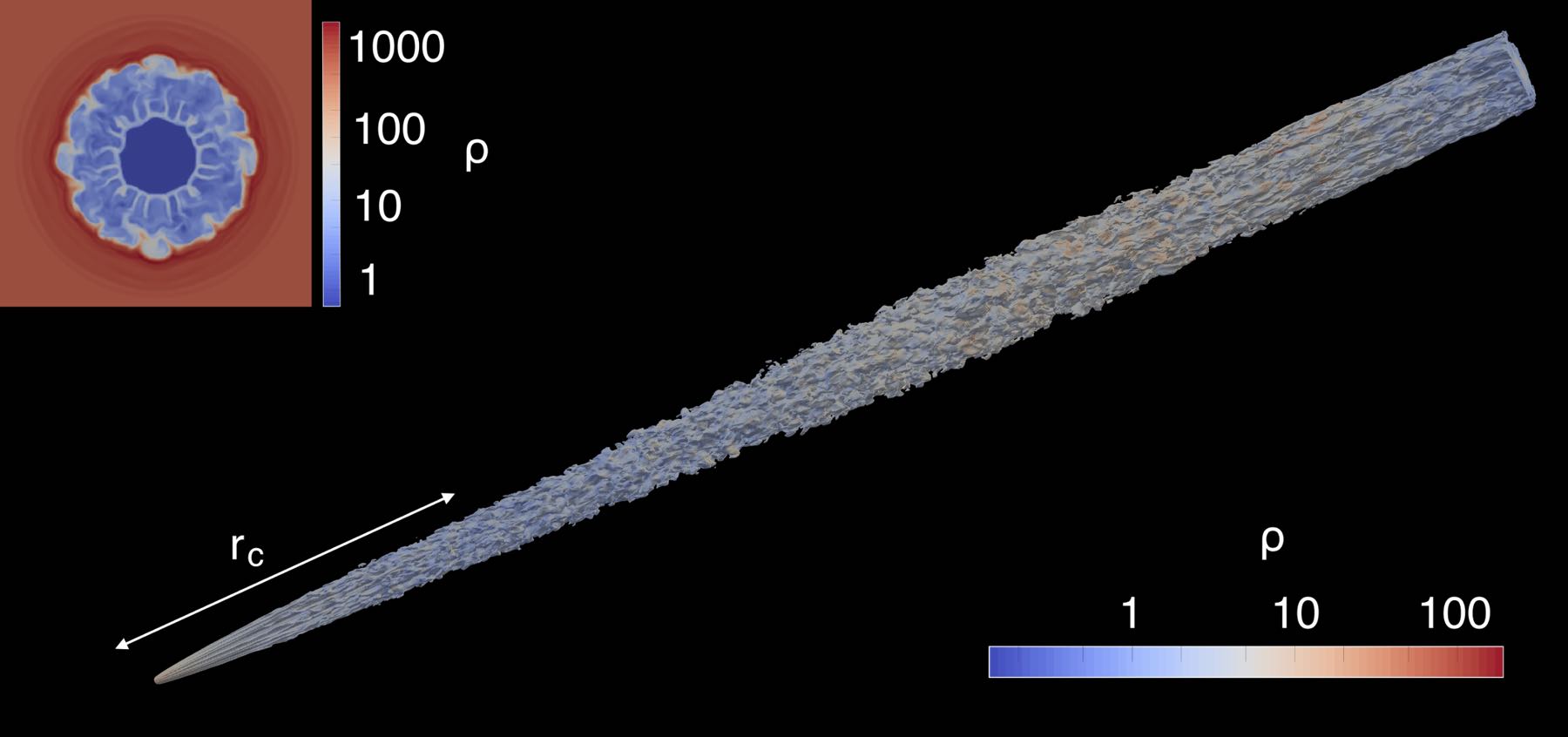}\\
\caption{{\bf 3D rendering of the model C5 at $t=13\,$kyr}.  The main image shows the distribution of the rest mass density on the  surface where the passive tracer $\tau=0.5$. The insert in the top-left corner shows the density distribution across the jet at  $z=r_{c}$.  Like in the models with larger opening angle, this jet also develops stream-aligned structures characteristic of the centrifugal instability.
}
\label{Figure:7}
\end{figure}

\end{document}